\documentclass
[aps,singlecolumn,prl,superscriptaddress,nobibnotes,showpacs,showkeys,10pt]{revtex4}%
\usepackage{amsfonts}
\usepackage{amsmath}
\usepackage{amssymb}
\usepackage{graphicx}%
\begin{document}
\title{Nonlinear wave-wave interactions in quantum plasmas}
\author{A. P. Misra}
\email{apmisra@visva-bharati.ac.in}
\affiliation{Department of Physics, Ume{\aa } University, SE--901 87 Ume{\aa }, Sweden.}
\affiliation{Department of Mathematics, Visva-Bharati University, Santiniketan-731 235, India.}
\author{P. K. Shukla}
\email{profshukla@yahoo.com}
\affiliation{Department of Physics, Ume{\aa } University, SE--901 87 Ume{\aa }, Sweden.}
\affiliation{RUB International Chair, International Centre for Advanced Studies in Physical
Sciences, Faculty of Physics \& Astronomy, Ruhr University Bochum, D-44780
Bochum, Germany.}

\begin{abstract}
Wave-wave interaction in plasmas is a topic of important research since the
16th century. The formation of Langmuir solitons through the coupling of
high-frequency (hf) Langmuir and low-frequency (lf) ion-acoustic waves, is one
of the most interesting features in the context of turbulence in modern plasma
physics. Moreover, quantum plasmas, which are ubiquitous in ultrasmall
electronic devices, micromechanical systems as well as in dense astrophysical
environments are a topic of current research. In the light of notable
interests in such quantum plasmas, we present here a theoretical investigation
on the nonlinear interaction of quantum Langmuir waves (QLWs) and quantum
ion-acoustic waves (QIAWs), which are governed by the one-dimensional quantum
Zakharov equations (QZEs). It is shown that a transition to spatiotemporal
chaos (STC) occurs when the length scale of excitation of linear modes is
larger than that of the most unstable ones. Such length scale is, however, to
be larger (compared to the classical one) in presence of the quantum tunneling
effect. \ The latter induces strong QIAW emission leading to the occurrence of
collision and fusion among the patterns at an earlier time than the classical
case. Moreover, numerical simulation of the QZEs reveals that many solitary
patterns can be excited and saturated through the modulational instability
(MI) of unstable harmonic modes. In a longer time, these solitons are seen to
appear in the state of STC due to strong QIAW emission as well as by the
collision and fusion in stochastic motion. The energy in the system is thus
strongly redistributed, which may switch on the onset of Langmuir turbulence
in quantum plasmas.

\end{abstract}
\date{07 September, 2010}
\startpage{1}
\endpage{102}
\maketitle

\section{{\protect\large Introduction}}

Quantum Zakharov equations (QZEs) \cite{QZEs} describe the nonlinear
interaction of high-frequency (hf) quantum Langmuir waves (QLWs) and the
low-frequency (lf) quantum ion-acoustic waves (QIAWs) in quantum plasmas. This
set of equations basically extends the classical Zakharov equations (CZEs)
\cite{CZEs} with higher-order dispersive terms associated with the Bohm
potential. Unlike CZEs, QZEs have been deduced using a quantum fluid
model\ under\ the similar quasineutral assumption and the multiple time scale
technique. \ Recently, much attention has been paid to investigate the
dynamics of QZEs \ in the context of e.g., the formation of Langmuir solitons
through modulational instability (MI) as well as Langmuir turbulence through
the process of chaos (See, e.g., Refs.
\cite{Variational-Haas,LieSymmetry,ExactSolution,Temporal,Spatiotemporal1}).
The statistical properties of the QZEs have been analyzed using kinetic
treatment and to show that the quantum coupling parameter ($H$) can be
responsible for reducing the MI growth rate \cite{MI-Marklund}. The latter is
, however, shown to be maximized in the limit of $H=0$. A variational approach
was conducted to study the quantum effects on localized solitary structures
\cite{Variational-Haas}. \ Moreover, some general periodic solution using Lie
point symmetries \cite{LieSymmetry}, some exact solutions \cite{ExactSolution}
as well as both temporal \cite{Temporal} and spatiotemporal dynamics
\cite{Spatiotemporal1} in the context of chaos and Langmuir turbulence \ of
the QZEs have been investigated in the recent past. \ Furthermore, a
comprehensive work on the dynamics of Langmuir wave packets in three spatial
dimensions \cite{3DQZEs} as well as some investigations on arrest of Langmuir
wave collapse \cite{WaveCollapse} by the quantum effects can be found in the
recent works.

Notice that when the wave electric field is strong such that it approaches the
decay instability threshold, the interaction of QLWs and the QIAWs is said to
be in `weak turbulence', and then QLWs are essentially scattered off QIAWs. On
the other hand, when the electric field intensity is so strong that the MI
threshold is exceeded, the interaction results in `strong turbulence' regime
in which QLWs are typically trapped by the density cavities associated with
the QIAWs. Such phenomena can frequently occur in plasmas. Moreover, the
transfer of energy to few stronger modes with small spatial scales can take
place due to a chaotic process, and this energy transfer can be faster when
the chaotic process in a subsystem of the QZEs is well developed
\cite{Temporal}.

In the present work, we will investigate the full QZEs numerically especially
when the wave number of modulation is small enough from its critical value.
The latter, in turn, excites many unstable harmonic modes which in a longer
time collide and fuse into fewer new incoherent patterns due to QIAW emission.
We will show that since the critical wave number of modulation (below which
the MI sets in) depends on the quantum coupling parameter $H,$ the length
scale of excitation for the transition from temporal chaos (TC) to
spatiotemporal chaos (STC) is to be larger than the classical case. Moreover,
lower the values of $H<1$, the higher is the number of unstable harmonic
modes. Furthermore, we will show that the solitary waves thus formed due to MI
will lose their strength after a long time through random collision and fusion
among the patterns under strong QIAW emission. This process becomes quicker
whenever the density correlation due to quantum fluctuation becomes strong
with higher values of the quantum parameter $H$. The STC state is then said to
emerge, and the energy of the system is thus redistributed to new incoherent
patterns as well as to few stronger modes with small length scales.

\section{{\protect\large Spatiotemporal evolution\ of QZEs}}

The nonlinear interaction of QLWs and QIAWs is described by the following
one-dimensional QZEs \cite{QZEs}.%

\begin{equation}
i{\frac{\partial E}{\partial t}}+{\frac{\partial^{2}E}{\partial x^{2}}}%
-H^{2}{\frac{\partial^{4}E}{\partial x^{4}}}=n\,E\,, \label{1}%
\end{equation}%
\begin{equation}
{\frac{\partial^{2}n}{\partial t^{2}}}-{\frac{\partial^{2}n}{\partial x^{2}}%
}+H^{2}{\frac{\partial^{4}n}{\partial x^{4}}}={\frac{\partial^{2}|E|^{2}%
}{\partial x^{2}}}\,, \label{2}%
\end{equation}
where $E=E(x,t)$ is the slowly varying wave envelope of the hf electric field
and $n=n(x,t)$ is the lf plasma density perturbation due to QIAW fluctuation.
\ Also, $H={\hbar\,\omega_{i}/}k_{B}\,T_{e},$which is associated with the Bohm
potential, represents the ratio of the ion plasmon energy to the electron
thermal energy. Here $\hbar$ is the scaled Planck's constant, $k_{B}$ is the
Boltzmann constant, $T_{e}$ is the electron temperature and $\omega
_{i(e)}=\sqrt{n_{0}e^{2}/m_{i(e)}\varepsilon_{0}}$ is the ion (electron)
plasma frequency with $n_{0}$ denoting the constant background density and
$m_{i(e)}$ the ion (electron) mass. The electric field $E$ is normalized by
$\sqrt{16m_{e}n_{0}k_{B}T_{e}/m_{i}\varepsilon_{0}}$ , the density $n$ by
$4m_{e}n_{0}/m_{i}$. Moreover, the space and time variables are rescaled by
$(\lambda_{e}/2)\sqrt{m_{i}/m_{e}}$ and $m_{i}/2m_{e}\omega_{e}$, where
$\lambda_{e}$ is the electron Debye length. \ \ Note that by disregarding the
term $\propto H^{2}$ in Eqs. (\ref{1}) and (\ref{2}), one recovers the
well-known CZEs \cite{CZEs}. The latter have been widely studied in the
context of solitons, chaos and Langmuir turbulence in many areas of plasma
physics (see, e.g., Refs. \cite{STC1,STC2,Spatiotemporal2}). It is thus of
natural interest to investigate the QZEs in the quantum realm, which may be
useful for understanding the onset of plasma wave turbulence at nanoscales in
both laboratory and astrophysical plasmas.

The growth rate of MI can be obtained from Eqs. (\ref{1}) and (\ref{2}) by
assuming the perturbations of the form exp$(ikx-i\omega t)$ from a spatially
homogeneous pump electric field $E_{0}$ as \cite{MI-Marklund}%

\begin{equation}
\Gamma=\frac{1}{\sqrt{2}}\left[  \digamma k^{2}\sqrt{\digamma^{2}+8|E_{0}%
|^{2}-2\digamma^{4}k^{2}(2-\digamma^{2}k^{2})}-\digamma^{2}k^{2}%
(1+\digamma^{2}k^{2})\right]  ^{1/2}, \label{3}%
\end{equation}
where $\digamma=1+H^{2}k^{2}$ and $k<\sqrt{2}E_{0}/\digamma.$ It can be shown
that the growth rate is maximum at $H=0$ and decreases for increasing values
of $H$ with cut-offs at lower wave numbers of modulation \cite{MI-Marklund}.

In order to solve numerically the Eqs. (\ref{1}) and (\ref{2}) we choose the
following initial condition \cite{Spatiotemporal1,Spatiotemporal2}%

\begin{equation}
E(x,0)=E_{0}\left[  1+\beta\cos(kx)\right]  ,\text{ }n(x,0)=-\sqrt{2}%
E_{0}k\beta\cos(kx), \label{4}%
\end{equation}
where $E_{0}$ is the amplitude of the pump Langmuir wave\ field and $\beta$ is
a constant of the order of $10^{-3}$ to emphasize that the perturbation is
very small. \ We use Runge-Kutta scheme with time step size $dt=0.0001$ and
consider the grid size $2048$ so that $x=0$ corresponds\ to the grid position
$1024$. The spatial derivatives are approximated with centered second-order
difference approximations. The results are presented in Figs. 1-6 after the
end of the simulation with $t=200$ and $E_{0}=2.$ \begin{figure}[ptb]
\begin{center}
\includegraphics[height=3in,width=3.0in]{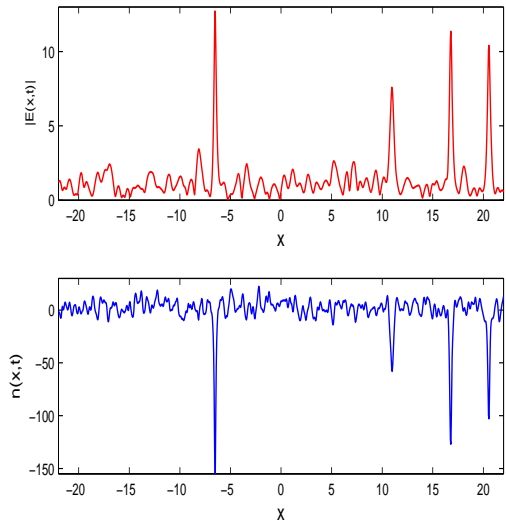}
\end{center}
\caption{(Color online) The profiles of the wave electric field $E(x,t)$
(upper panel) and the associated density fluctuation $n(x,t)$ (lower panel)
with respect to the space $x$ after time $t=200$ in the numerical simulation
of the QZEs for $k=0.14$, $H=0$ and $E_{0}=2$.}%
\end{figure}\begin{figure}[ptb]
\begin{center}
\includegraphics[height=3in,width=3.0in]{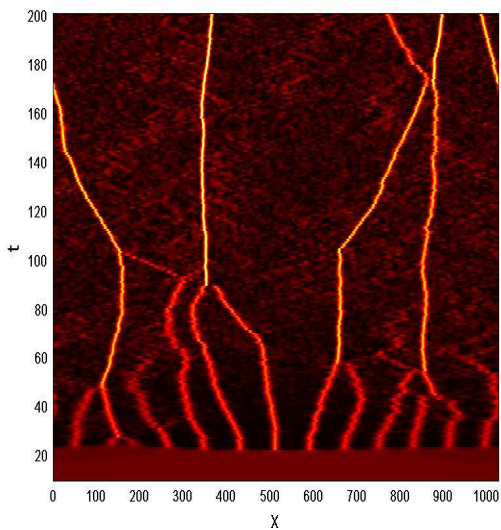}
\end{center}
\caption{(Color online) Space-time contour plots of \ $\left\vert
E(x,t)\right\vert =$constant corresponding to the evolution as in Fig. 1. This
shows that pattern selection leads to many solitary patterns which due to QIAW
emission collide and get fused into few incoherent patterns after a long time.
The STC state emerges. }%
\end{figure}Note that as above the MI sets in for wave numbers satisfying
$0<k<k_{c},$ where $k_{c}$ is a real root of the cubic $H^{2}k^{3}+k-\sqrt
{2}E_{0}=0,$ and\ $k=k_{c}$ defines the curve along which pitchfork
bifurcation takes place. Moreover, the dynamics is subsonic in the regime
$k_{c}/2<k<k_{c}$ where the MI growth rate is small. As $k$ is lowered from
$k_{c}/2,$ many unstable modes with higher harmonic modes will be excited.
\ Again, the master mode can, in principle, result in the excitation of $N-1$
unstable harmonic modes where $N=[\bar{k}^{-1}]$ with $\bar{k}=k/\sqrt{2}%
E_{0}.$ There may also exist many solitary patterns with spatially modulated
length $l_{m}=L/m,$ where $m=1$ is for master mode ($l_{1}=L=2\pi/k$) and
$m=2,3,....,M$ are for the unstable harmonic modes. As a result, the envelope
$E$ can be expressed as: $\allowbreak$%

\begin{equation}
E(x,t)=%
{\displaystyle\sum\limits_{m=1}^{M}}
E_{m}(t)\exp(im\bar{k}x)+%
{\displaystyle\sum\limits_{m=M+1}^{\infty}}
E_{m}(t)\exp(im\bar{k}x), \label{5}%
\end{equation}
in which the first term on the right-side of Eq. (\ref{5}) comes from the
master mode and unstable harmonic modes with $M<N-1$ being due to pattern
selection, whereas the second term is due to the nonlinear interaction among
the patterns formed. \begin{figure}[ptb]
\begin{center}
\includegraphics[height=3in,width=3.0in]{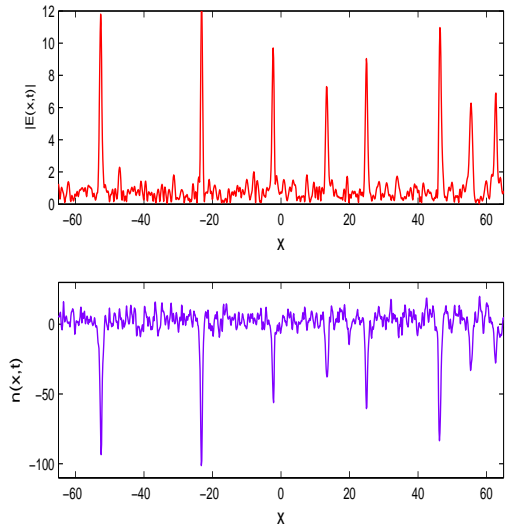}
\end{center}
\caption{(Color online) The same as in Fig. 1, but for $H=0.2$, $k=0.048$ and
other parameters remain the same as in Fig. 1.}%
\end{figure}\begin{figure}[ptb]
\begin{center}
\includegraphics[height=3in,width=3.0in]{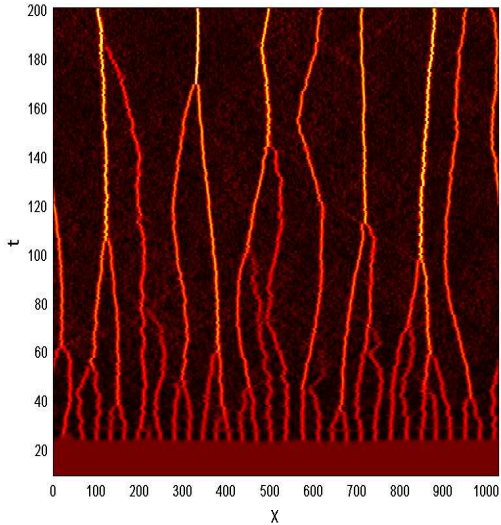}
\end{center}
\caption{(Color online) Contours of \ $\left\vert E(x,t)\right\vert =$constant
with respect to space and time corresponding to the evolution as in Fig. 3.
This shows that pattern selection leads to many more solitary patterns than
the classical case (Fig. 2) which due to QIAW emission collide and get fused
into few incoherent patterns after a long time. The collision is random and
not confined between two patterns. Few stable modes are also excited. The
system energy is then redistributed to new incoherent patterns and few stable
modes with small length scales. The system is in the state of STC.}%
\end{figure}

The profiles of the electric field and the density fluctuation at the end of
simulation are shown in Fig.1 for $k=0.14$ and $H=0$. We observe an excited
electric field of the order $|E|\sim12$ highly correlated with density
depletion $n\sim-152$ or $n/n_{0}\sim-0.33.$ From Fig. 2, we find that many
solitary patterns are formed from the master mode and unstable harmonic modes
by means of pattern selection. Two solitary patterns initially peaked in
$100<x<200$ collide to form a new mode which again collides with the master
pattern peaked at $x\sim50$ and fuses into a new strong mode. After few more
other collisions and fusions, there remain only four distorted patterns. The
system is then said to emerge the STC state. When $k$ is further lowered many
solitary trains are excited and saturated. As soon as the collisions and
fusions among the solitary patterns take place and the new incoherent trains
are formed under strong QIAW emission, these incoherent patterns then collide
with some others repeatedly and finally there remain few incoherent patterns
in stochastic motion with the greater part of the system energy. Meanwhile
many stronger modes are also excited to share the energy. These are shown in
Figs. 3, 4 and Figs. 5, 6 for different values of $H$: $H=0.2$ and $0.5$
respectively. We observe that though the number of remaining modes are the
same for both the cases, however, many more (compared to Fig. 6) solitary
patterns are seen to form initially (Fig. 4) which later participate into the
collisions and fusions under QIAW emission. Moreover, the new incoherent
patterns formed with the lapse of long time are more stronger as in Fig. 6
than those in Fig. 4. Furthermore, for higher values of $H$ (Fig. 6) the
collisions and fusions take place at an earlier time than the case of lower
$H$ (Fig.4). Thus, a certain amount of energy which was initially distributed
among many solitary waves, will now be transferred to a few incoherent
patterns as well as to some stable higher harmonic modes with short wave
lengths. So, if initially there form many solitary pattern trains due to MI
with different modulational lengths, collision and fusion among most them can
lead to the STC state. There must exist a critical wavelength or wave number
$k,$ at which the transition from temporal chaos to STC takes place. This
value, in the quantum case ($H\neq0$) is quite different from the classical
one ($H=0$). For detail investigations readers are referred to works, e.g., in
Refs. \cite{Spatiotemporal1,Spatiotemporal2}. \begin{figure}[ptb]
\begin{center}
\includegraphics[height=3in,width=3.0in]{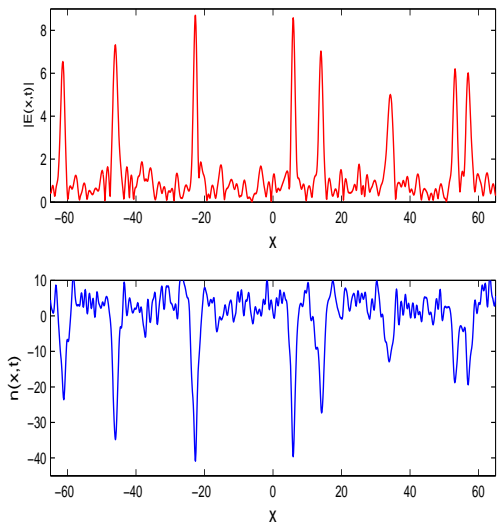}
\end{center}
\caption{(Color online) The same as in Fig. 3, but for different $H=0.5$.
Other parameters remain the same as in Fig. 3. }%
\end{figure}\begin{figure}[ptb]
\begin{center}
\includegraphics[height=3in,width=3.0in]{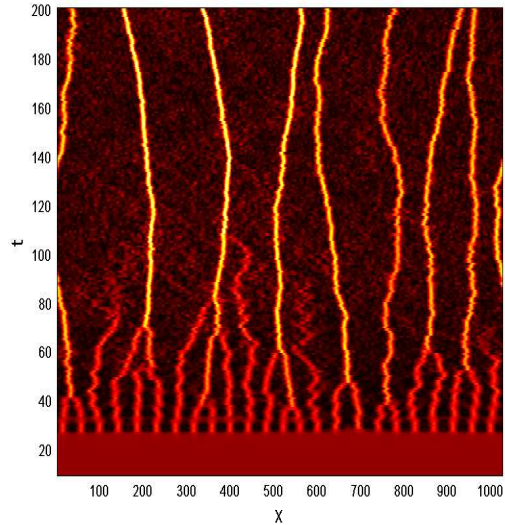}
\end{center}
\caption{(Color online) Contours of \ $\left\vert E(x,t)\right\vert =$constant
with respect to space and time corresponding to the evolution as in Fig. 5.
Here the pattern selection leads to many solitary patterns, but less than that
in Fig. 4. Though, the collision is random and not confined between two
patterns, but they fuse into new incoherent patterns in a shorter time than
the case of $H=0.2$ (Fig. 4) due to strong QIAW emission. The system is also
in the STC state.}%
\end{figure}

\section{Conclusion}

We have performed a simulation study of the QZEs to show that many coherent
solitary patterns can be excited and saturated through the MI of unstable
harmonic modes by a modulation wave number of QLWs. It is observed that there
exist critical values of $k$ for which the motion of the coherent solitary
patterns is in the state of TC or in STC. The transition from TC to STC occurs
when $k\lesssim0.14$ for $H=0$ or, for $k\lesssim0.048\ $when $H\gtrsim0.2.$
It is shown that the dispersion due to quantum tunneling induces strong QIAW
emission leading to collision and fusion among the patterns to occur at an
earlier time than the classical case. The Collision and fusion among some
trains take place and the new incoherent pattern trains are formed
accompanying strong QIAW emission due to quantum effects. The STC state is
then said to emerge. As a result, the system energy in the STC state is
spatially redistributed in the process of pattern collision, fusion and
distortion, which may switch on the onset of Langmuir turbulence in quantum plasmas.

\section{Acknowledgments}

This work was prepared in honor of Professor P. K. Shukla's 60th birthday.
\ A. P. M gratefully acknowledges support from the Kempe Foundations, Sweden.

\end{document}